\documentclass[fleqn,twoside]{article}
\usepackage[headings]{espcrc2}
\usepackage{graphicx}
\usepackage[figuresright]{rotating}

\newcommand{\AmS}{{\protect\the\textfont2
  A\kern-.1667em\lower.5ex\hbox{M}\kern-.125emS}}

\title{Quantum chaos, localized states and clustering in excited
 spectra of  Jahn-Teller models}

\author{Eva Majern\'{\i}kov\'a
\address[BA]{ Institute of Physics, Slovak Academy of
Sciences, D\'ubravsk\'a cesta 9, SK-84 511 Bratislava, Slovak
Republic}$^,$
\address[OL]{
 Department of Theoretical Physics, Palack\'y University, T\v r.
17. listopadu 50, CZ-77207 Olomouc, Czech Republic} S. Shpyrko
\addressmark[OL]}

\runtitle{Quantum chaos in Jahn-Teller models} \runauthor{Eva
Majern\'{\i}kov\'a, S. Shpyrko}

\begin{document}

\begin{abstract}
We studied complex spectra of spin-two boson systems represented
by E$\otimes$e and E$\otimes (b_1+b_2)$ Jahn-Teller models. For
E$\otimes$e, at particular rotation quantum numbers we found a
coexistence of up to three regions of the spectra, (i) the
dimerized region of long-range ordered (extended) pairs of
oscillating levels, (ii) the short-range ordered (localized) "kink
lattice" of avoiding levels, and (iii) the intermediate region of
kink nucleation with variable range of ordering. This structure
appears above certain critical line as a function of interaction
strength. The level clustering and level avoiding generic patterns
reflect themselves in several intermittent regions between up-to
three branches of spectral entropies. Linear scaling behavior of
the widths of curvature probability distributions provides the
conventionally adopted indication for the presence of quantum
chaos. The mapping onto classical integrable Calogero-Moser gas
provided useful insight into the complex level dynamics, including
the soliton collisions representing the level avoidings and, in a
range of model parameters, a novel view on the notion of quantum
chaos formulated in terms of quantum numbers via the logistic
equation.  We found that apart from two limiting cases of
$E\otimes(b_1+b_2)$ model ($E\otimes e$ and Holstein model) the
distribution of nearest neighbor spacings of this model is rather
stable as to the change of parameters and different from Wigner
one. This limiting distribution assumably shows scaling
$\sim\sqrt{S}$ at small $S$ and resembles the semi-Poisson law
$P(S)= 4S \exp (-2 S)$ at $S\geq 1$. The latter is believed to be
universal and characteristic, e.g. at the transition between metal
and insulator phases.
 \vspace{1pc}

\end{abstract}

\maketitle

\section{Introduction}
\label{intro}

The results of intensive research in quantum mechanics show
 (\cite{Gutzwiller:1990}-\cite{Nakamura:1993}) that the smooth wavelike
 quantum world contains elements of quantum complexity and that symptoms of chaos enter even into
 the wave patterns associated with atomic energy levels.
 Starting with nuclear physics \cite{Eckhardt:1988}, one observes complexity in quantum
 phenomena in spectral features,
 eigenfunctions, dynamics of molecular processes, solid state
 physics and information science (Berry \cite{Berry:1987} suggested the "quantum chaology" being
 the  adequate term for  a quantum system whose classical counterpart exhibits chaos
 instead the mostly used and widely accepted term "quantum
 chaos").\\
Besides the quantum systems that exhibit chaos in the classical or
a semiclassical limit there exists a class of quantum systems that
do not have a naive classical or semiclassical counterpart. In
addition, the way of passing to semiclassical approximation, e.g.,
in spin-boson systems is not unique and presents essential
ambiguity due to different possible ways of decoupling quantum
variables\cite{Graham:1984}.
 Different ways of performing
semiclassical approximation are known to lead to different answers
concerning the chaotic behaviour of the system. This ambiguity
means that a classical analogue of such a model is not
well-defined and
cannot be a reliable object in exploring the quantum chaos issues.\\
 The class of "spin-boson"
 models, e.g., two and more levels coupled to bosons (vibrons, phonons,
 photons) were reported to exhibit chaotic phenomena first by
 Lewenkopf et al \cite{Lewenkopf:1991} and Cibils et al \cite{Cibils:1995},
 and many others. Fujisaki et al (\cite{Fujisaki:2001,Fujisaki:2004}) investigated
 the systems with the absence of a reasonable
 quasiclassical limit.  There such phenomena as quantum tunneling and
multi-mode nonadiabatic fluctuations lead to complicated wave packet
dynamics which breaks down the Born-Oppenheimer adiabatic
approximation.\\
Typical representative of such a non-adiabatic system is the
two-level molecular system coupled with two vibron modes of
different symmetry against the transformation of reflection
 --  the Jahn-Teller (JT) class of models.
 The chaos in JT molecules was investigated within a generalized model
 by Cederbaum et al in a series of papers \cite{Cederbaum:1983}.
Recently, Yamasaki et al \cite{Nakamura:2003} for the first time
investigated a possibility of chaos in spectra of the E$\otimes$e
JT model.  Their analysis was based on the approximation of the
Hamiltonian by separating it into the
 adiabatic "Mexican hat" potential and an additional
part supplemented via a parameter for a nonlinear mode coupling of
a trigonal symmetry to include the effect of fluctuations and
non-integrability. Thus, the nonlinearity was included via mode
coupling in addition to the mean field bare part of the
Hamiltonian. The authors concluded that the quantum chaos
reflected itself in the Wigner level spacing distribution as a
consequence of the said nonlinearity of the Hamiltonian, meanwhile
for the linear part the absence of these patterns was stated.\\
In the light of said problems due to the absence of a reasonable
quasiclassical approximation we have used an exact numerical
approach to the characteristics of excited (quasi-continuum)
spectra of the E$\otimes$e JT model
\cite{Majernikova:2006a,Majernikova:2006b}. This approach differs
from that by Yamasaki et al. in the following: a) we {\it do not}
introduce an explicit nonlinearity into the initial Hamiltonian.
However, its SU(2) symmetry involves an {\it intrinsic
nonlinearity} which is revealed by the exact elimination of the
electronic degrees of freedom \cite{Majernikova:2003}.
 b) We started from the E$\otimes (b_1+b_2)$ model
 with different coupling strengths for both modes.
 The rotation symmetric E$\otimes$e model represents its
particular case with equal coupling constants and thus with the
symmetry of higher (rotational) symmetry group [the difference of
the coupling constants in realistic systems is likely to be
caused, for example, by the spatial anisotropy of crystals]. The
importance of starting from the more general symmetry is evident
from the nature of quantum fluctuations: namely, the variational
approach to E$\otimes (b_1+b_2)$ model used for the calculation of
the ground state \cite{Majernikova:2003} was shown to yield the
largest deviations just for the rotation symmetric case. In the
ground state an abrupt change of the energy at equal couplings is
an artefact of the adiabatic approximation: the energy region of
the quantum E$\otimes $e model is situated
 within a smeared border around $\beta/\alpha=1$ ($\alpha$ and
 $\beta$ being the coupling strengths of the antisymmetric and the symmetric
 modes, respectively)
between the "selftrapping" ($\alpha>\beta$) and "tunneling"
($\alpha <\beta$) part of the ground state of the model with
broken rotational symmetry. In the smeared transition region both
phases coexist and are quantum correlated (entangled) via
nonlinear correlations between the modes. There the
phonon-assisted (by the symmetric mode) tunneling contribution to
the energy [from the admixture of two reflection symmetric levels
in the excitation reflection Ansatz for the variational wave
function \cite{Majernikova:2003}] results in the essential
decrease of the ground state energy because of changing parity of
the wave function by the parity reversal of the phonon
$1$-coordinates by the operator $R$, $R b_1=-b_1R$, $ \
\beta\langle \Psi_ 0^+ |\hat Q_2 R | {\Psi_1}^- \rangle
=\beta\langle \Psi_ 0^+|\hat Q_2| \Psi_1^+\rangle \gg \beta\langle
\Psi_ 0^+ |\hat Q_2 | {\Psi_1}^-\rangle .$  Therefore, it has its
consequence in a dramatic increase of the Ham factor
and in the corresponding decrease of the energy.\\
Excited states follow the structure of dominating selftrapping
($\alpha$) or
 tunneling ($\beta$) interactions. The mixing of the phases
due to the tunneling is amplified by the analogous to the above
matrix element via the reflection change of the parity  in matrix
elements with excited wave functions. Respective decrease of the
energy manifests itself
in the avoiding of the excited energy levels.\\
The phase transition at $\alpha=\beta $ to the rotational symmetry
phase provides a representation by the additional rotational
quantum number $j$. At given $j$ in E$\otimes $e model the
analogous phonon assisted tunneling between adjacent excited
levels occurs and is responsible for the flip (kink) between the
levels (this mechanism can be well recognized, e.g., in Fig. 1b).
This nonlinear effect causes a mixture of the adjacent (even and
odd) levels at stronger couplings and leads to the nucleation of
the new
 kink lattice phase with increasing coupling constant.
 The tunneling mechanism due to short-range quantum fluctuations at
large $j$ is then the origin of peculiarly irregular behaviour of
the related level spacing distributions and does not allow for
full developing of the Wigner chaos.
  The phonon assisted tunneling and nucleation of the kink lattice phase
can be also demonstrated  within the formalism of a Calogero-Moser
gas of pseudo-particles with repulsive interaction
\cite{Pechukas:1983}.
\section{Hamiltonian approach: two-level-two-boson Hamiltonians }
The aim of this Section is to explore the numerical solution of
the E$\otimes $e Hamiltonian for the excited energy spectrum and
wave functions in search for the statistical evidences of the
symptoms of chaotic behaviour at the quantum level. Namely, the
participation ratios, spectral entropies and widths of level
curvature
distributions are anticipated to provide the relevant indications thereof.\\
The E$\otimes $e Hamiltonian with the parameter of
nonintegrability $\alpha$ is written as
\begin{equation}
H=  (b_{1}^{\dag}b_{1} +b_{2}^{\dag}b_{2}+1 )I + \alpha
(b_{1}^{\dag}+b_{1})\sigma_{z}
 -\alpha (b_{2}^{\dag}+b_{2})\sigma_{x},
 \label{1}
\end{equation}
 where $\sigma_{j}$  are Pauli matrices, $I$ is the unit matrix.
The pseudospin notation is used for the representation of a two-level one-electron (spinless) system.\\
For the secular equations of the vibronic parts of the {\it gerade}
solution
 $K=+1$ ($j-1/2=0,2,4,6,\dots; -2,-4,\dots$)  the
 representation of  radial coordinates
\begin{eqnarray}
 H_r f_i+\frac{1}{2 r^2}\left(j+\frac{1}{2}(-1)^i \right)^2 f_i+\alpha\sqrt{2}r
f_k=E f_i , \nonumber\\
 i\neq k, \ i, k=1,2  \end{eqnarray}
 \begin{equation}H_r=
-\frac{1}{2r}\frac{\partial}{\partial r} \left(
r\frac{\partial}{\partial r}\right) +\frac{1}{2}r^2
\end{equation}
will be used, and radial wave functions are assumed in the form
$f=\sum c_n\Phi_n $, where
\begin{equation}
\Phi_n = \left\{\matrix {\Phi^+(n_r,j,r), \hfill \quad n=2n_r\,,\cr
\Phi^-(n_r,j,r), \quad  n=2n_r+1\,. }\right.
\end{equation}
 The only non-zero off-diagonal elements are
\begin{eqnarray}
f_{n_r n_r}\equiv\sqrt{2}\int\Phi^+_{n_r}\Phi^-_{n_r} r\cdot r dr=
\sqrt{2}\sqrt{n_r+1+ \left| j-\frac{1}{2} \right|}\,,
\label{f:nn}\\
f_{n_{r}+1 n_r}\equiv\sqrt{2}\int\Phi^+_{n_r+1}\Phi^-_{n_r} r\cdot
r dr=-\sqrt{2}\sqrt{n_r+1}\,. \label{f:nnn}
\end{eqnarray}
The secular equations are written in the usual tridiagonal form
($E^{(0)}$ is the "unperturbed" energy for corresponding $n,j$),
\begin{eqnarray}
c_nE^{(0)}_n+\alpha(c_{n+1}f_{n_r n_r}+c_{n-1}f_{n_r n_r-1}) =E
c_n, \nonumber\\  n=2n_r\,, \nonumber\\
c_nE^{(0)}_n+\alpha(c_{n-1}f_{n_r n_r}+c_{n+1}f_{n_r+1 n_r}) =E
c_n, \nonumber\\
 n=2n_r+1 \, .
 \label{28}
\end{eqnarray}
Numerical solutions of (\ref{28}) for the spectra and
corresponding wave functions are exemplified in Figs. (1-4).
\begin{figure}
\includegraphics[width=0.8\hsize]{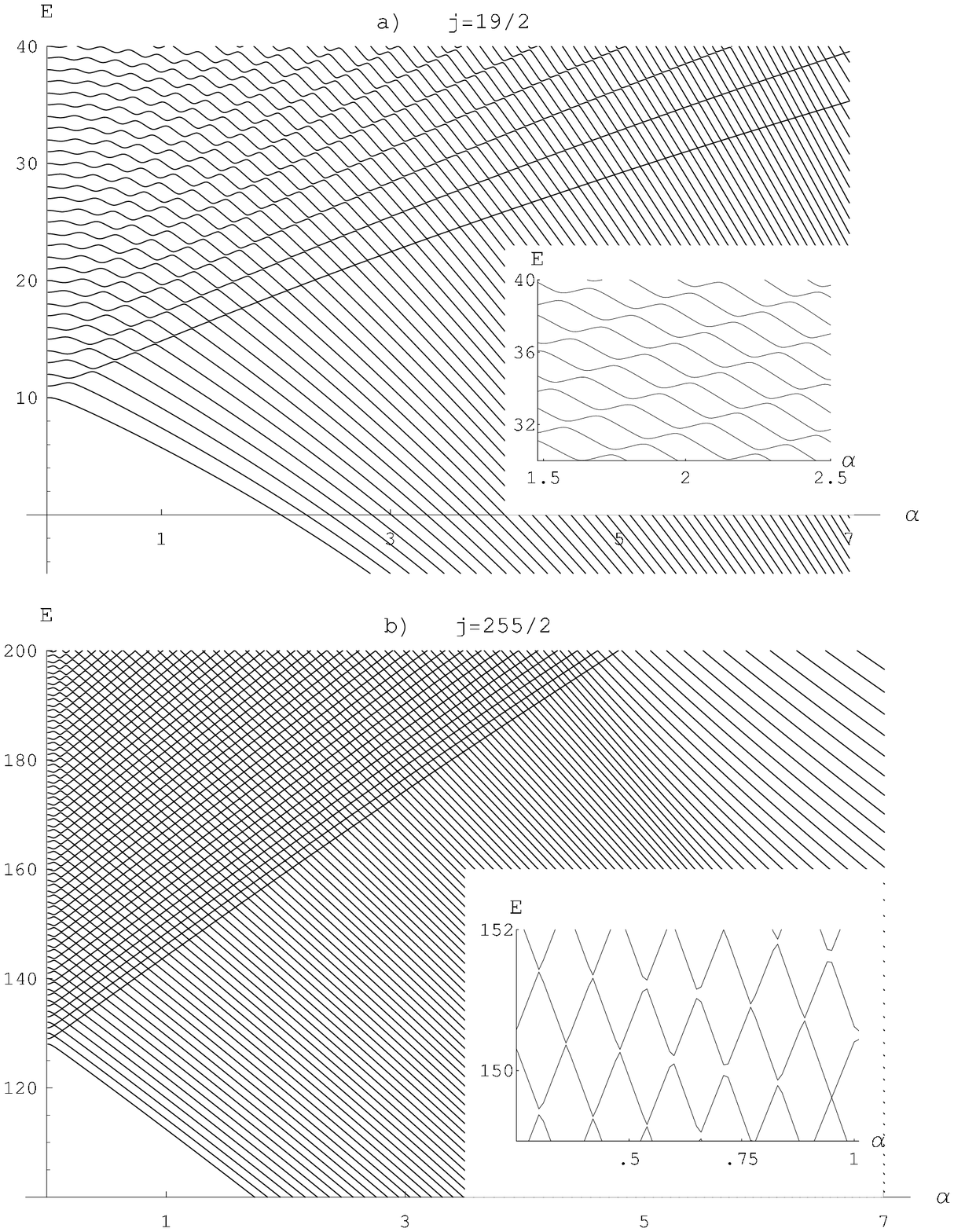}
\caption{Energy spectra for $j=19/2,255/2$ as function of $\alpha$.}
\end{figure}
\begin{figure}
  \includegraphics[width=0.9\hsize]
{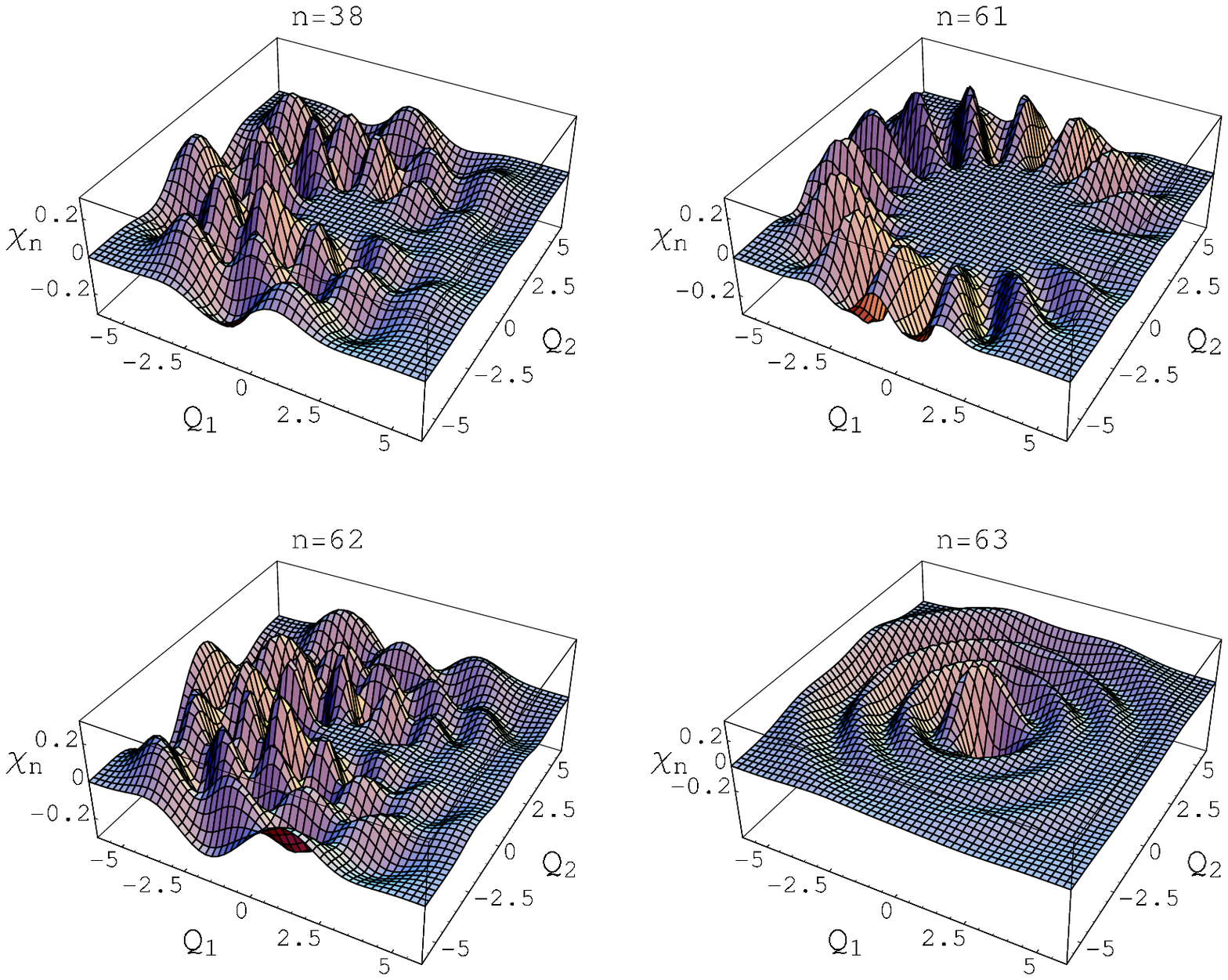}
\caption{Numerical wavefunctions $\chi_n$ in the
plane $Q_1\times Q_2$, $\alpha=\beta=2$. States n=61, 63 are
localized ("exotic"). }
\end{figure}%
\begin{figure}
\includegraphics[width=0.9\hsize]{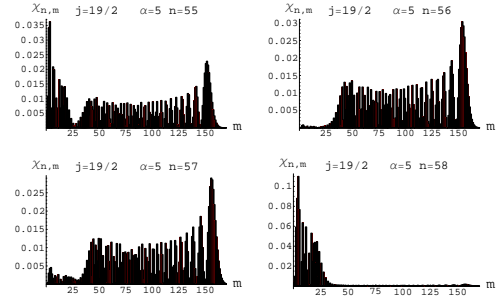}
\caption{Projections of exact symmetric ($\beta=\alpha$) states
$\chi_n$ on the harmonic oscillator base,
$\chi_{n,m}=|\langle\chi_n|\Phi^0_m\rangle|^2$. }
\end{figure}
\begin{figure}
\includegraphics[width=0.9\hsize]{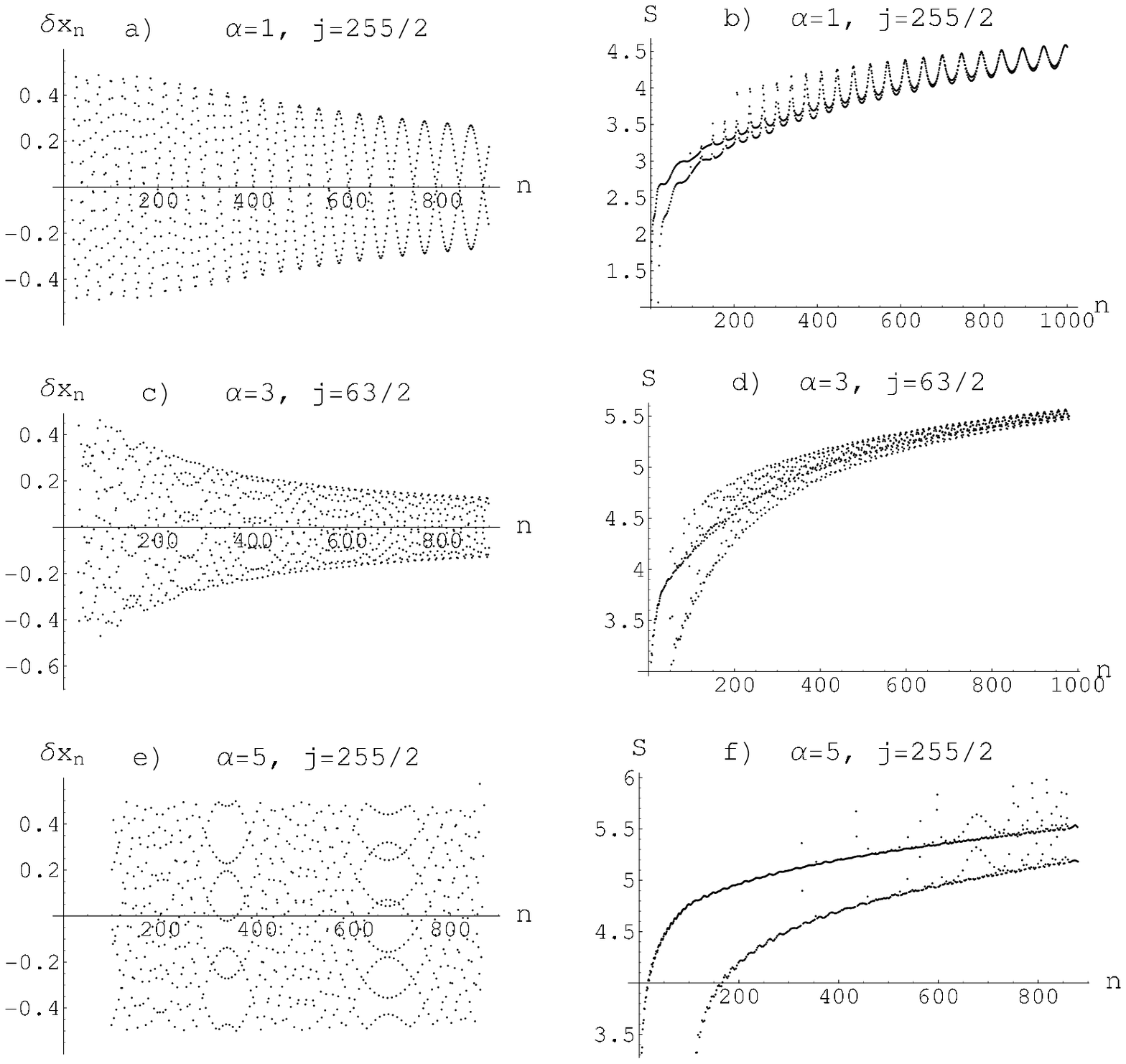}
\label{figW5} \caption{Reduced  energies and spectral entropies of
wavefunctions $\chi_n$ for different couplings $\alpha $ and $j$.
}
\end{figure}

\section{Statistical methods: Level spacing and level curvature probability distributions,
long-range statistics} The multitude of avoided crossings of
energy levels is generally claimed to be a testimony of quantum
chaos. The fundamental problem of the quantum-mechanical
manifestation of classical chaotic motion was investigated by
statistical methods based on the random matrix theory (RMT). It
setforth reference patterns to which the presumably quantum
chaotic systems should conform \cite{Eckhardt:1988,Mehta:1960}.
The conventionally adopted statistical measures are either the
short- or long-range statistics. The common representative of the
former is the distribution of nearest neighbour level spacings
(NNS), as well as the statistics involving the "curvatures"
$K\equiv d^2E_n/d\alpha^2$ as functions of the nonintegrability
parameter $\alpha$. Among the latter a generally adopted tool is
the $\Delta_3$-statistics \cite{Mehta:1960}.
 Both short- and long-range statistical
measures have reference formulas provided within the frames of the
RMT. Thus, for the distribution of NNS the prediction for the
fully developed chaos is the Wigner surmise $P(S) \sim S
\exp(-S^2)$, for the opposite situation of the regular NNS
distribution the Poisson distribution is representative
$P(S)=\exp(-S)$ (in the above formulas the spectra are supposed to
be scaled so that $\langle S \rangle = 1$ locally). The RMT leads
to the indication of universality of level fluctuations
conditioned solely by the time-reversal symmetry of the underlying
Hamiltonians: the fluctuation patterns show a transition from the
Poisson type to the Gaussian (orthogonal, unitary, symplectic)
type as the corresponding classical system shifts from the
integrable to the chaotic regime.

\subsection{Spin one-boson model }
The class of non-integrable spin-boson models is known to exhibit
quantum chaotic symptoms for one-boson many-level ($2j+1, j=1/2,
3/2,...$) cases. Respective Hamiltonian
\begin{eqnarray}
\hat H&= & \omega  b^{\dag} b  + \omega_0 \sigma_z + \lambda [
\sigma_+b +\sigma_-b^{\dag} + \nonumber \\
& & \epsilon (\sigma_+b^{\dag}+\sigma_- b)]\,, \nonumber \\
  \sigma_{\pm}&=&   \sigma_x\pm\sigma_y
 \end{eqnarray}
 is parametrized by the nonintegrability parameter $\epsilon$.
 For $\epsilon=1$ we obtain the exciton (Holstein) model (non-integrable, with evidence
 for chaos \cite{{Graham:1984}}-\cite{Cibils:1995}), while for $\epsilon=0$ one
 arrives at the
Jaynes-Cummings model which is integrable.
\subsection{ E$\otimes(b_1+b_2)$ and E$\otimes$e Jahn-Teller Model }
If $\alpha\neq \beta$ but $|\alpha-\beta|\ll \alpha$  in
(\ref{1}), there appear correlations between the excited levels
with $j\neq j^{'}$ of the Hamiltonian in radial representation due
to the term \cite{Majernikova:2006a}
\begin{equation}
(\alpha-\beta)\sqrt{2}r\cdot \left(\matrix{\sin\phi R_{ph} \ & 0 \cr
0 \ & -\sin\phi R_{ph}}   \right) \, .
\end{equation}
Respective additional avoided level crossings change dramatically
the probability distributions of level spacings when compared to
those of the $E\otimes e$ model ($\alpha=\beta$), Fig. 5.
\begin{figure}
\includegraphics[width=1.\hsize]
{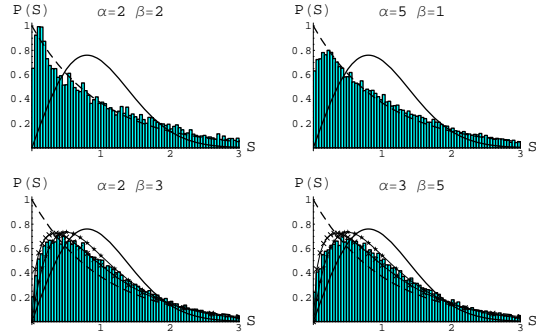}
\caption{Sample nearest-neighbor distributions of
levels (unfolded and scaled to $\langle S \rangle = 1 $) for
different values of $\alpha$, $\beta$. The curves on the histograms
represent Poisson (long dashed), Wigner (full), semi-Poisson (stars)
and $\sqrt{S}$ (crosses) distributions.}
\end{figure}

 Fig. 5 presents the typical forms of the NNS distributions for
different $\alpha, \beta$ of the JT model with broken radial
symmetry. It is seen that for the special cases of higher symmetry
the NNS distributions indeed are close to the Poisson distribution
(first row). Meanwhile far from the cases $\alpha=\beta$ and
$\alpha \ll \beta$, $\alpha \gg \beta$ the NNS distributions do
not show a tendency to the RMT prediction (Wigner surmise). The
well defined limit for these distributions seems not to have an
analogue in the present versions of the RMT ensembles and presents
some intermediate statistics about which we can make now only
plausible assumptions. Thus, it is found that the semi-Poisson
distribution $P(S)=4S \exp(-S)$ \cite{Evangelou:00,Shklovskii:93}
describes well this intermediate statistics in the domain of large
$S\geq 1 $ although for small $S$ the limiting distribution seems
to scale as $P(S)\sim S^\nu$ with $\nu$ ranging between $0.3\div
0.5$. The semi-Poisson statistics has recently found its
application as a limiting case of level statistics at the M-I
transition in Anderson localization models and, therefore, such an
analogy seems to be a promising item. As the second fitting
formula (see Fig. 5, second row) we use the
$\sqrt{S}$-distribution $P(S)\sim \sqrt{S} \exp{(-3/2 S)}$ which
conforms to the interpolation of Brody type.
\begin{figure}
\includegraphics[width=0.8\hsize]{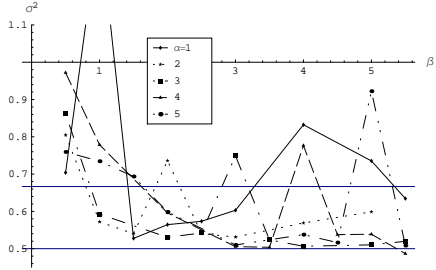}
\caption{Standard deviations $\sigma^2=\langle S^2 \rangle - 1$ of
the NNS distributions. Grid lines correspond to the semi-Poisson
(0.5) and $\sqrt{S}$ (2/3) values. Far from the higher symmetry
cases (Holstein and E$\otimes$e JT model) the dispersion tends to
a {\it well defined limit} $\sigma^2\simeq 0.5$. The fully chaotic
Wigner NNS distribution would have $\sigma^2\simeq 0.28$.}
\end{figure}
From the numerical analysis of the nearest-neighbour  level spacing
probability distributions  $P(S)$ we conclude the diagnosis of the
quantum chaotic patterns of JT models: \\

\begin{enumerate}
\item  For small $S$, $P(S)\sim S^{\nu}$, where $\nu $ acquires a
non-universal value
 from the interval ($0.3, 0.5$).
\item Except for small $S$, the limit cases of $E\otimes
(b_1+b_2)$ model of higher symmetry  (E$\otimes$e and Holstein)
show up the close-to-Poisson distribution of NNS, especially in
tails (Fig. 5, first row).
\item Far from the higher symmetry cases the NNS distribution
tends to a universal law which is however markedly distinct from
the Wigner distribution expected for the fully developed quantum
chaos (Fig. 5, second row).
\item The dispersion tends to a {\it well defined and universal
limit} $\sigma^2\simeq 0.5$ (Fig. 6) referring to the semi-Poisson
distribution $P(S)= 4S \exp(-2S)$. This distribution is known as a
critical distribution at the M-I transition in Anderson
localization models.
\end{enumerate}

For the $E\otimes e$ model a more pronounced indication for the
presence of quantum chaos than the NNS level spacing distribution
is the conventionally adopted \cite{Nakamura:1985} linear scaling
behaviour of the widths of curvature probability distributions
$\Delta K\equiv \langle \Delta^2 K\rangle $ as a function of
nonintegrability parameter ($\alpha $),
  Fig. 7.
The quantum number $j$ is a measure of a departure from the
semiclassical limit $j\rightarrow \infty$. From
 Fig. 8 it is seen that the widths of the curvature distribution show
 a scaling $\Delta K \sim j^\mu \alpha^\nu$ with $\mu\simeq 1$ and
 $\nu\simeq 0.5$. The latter observation, as well as the form of
 the typical profiles of $P(K)$ allows for the
 interpretation of the evolution of $P(K)$ in the pseudotime
 $\alpha$ as a solution of an ordinary diffusion-like equation with added
 ballistic ("telegraph") term $ d^2 P(K)/d\alpha^2$  \cite{Uchaikin:03}.
\begin{figure}
\includegraphics[width=0.7\hsize]{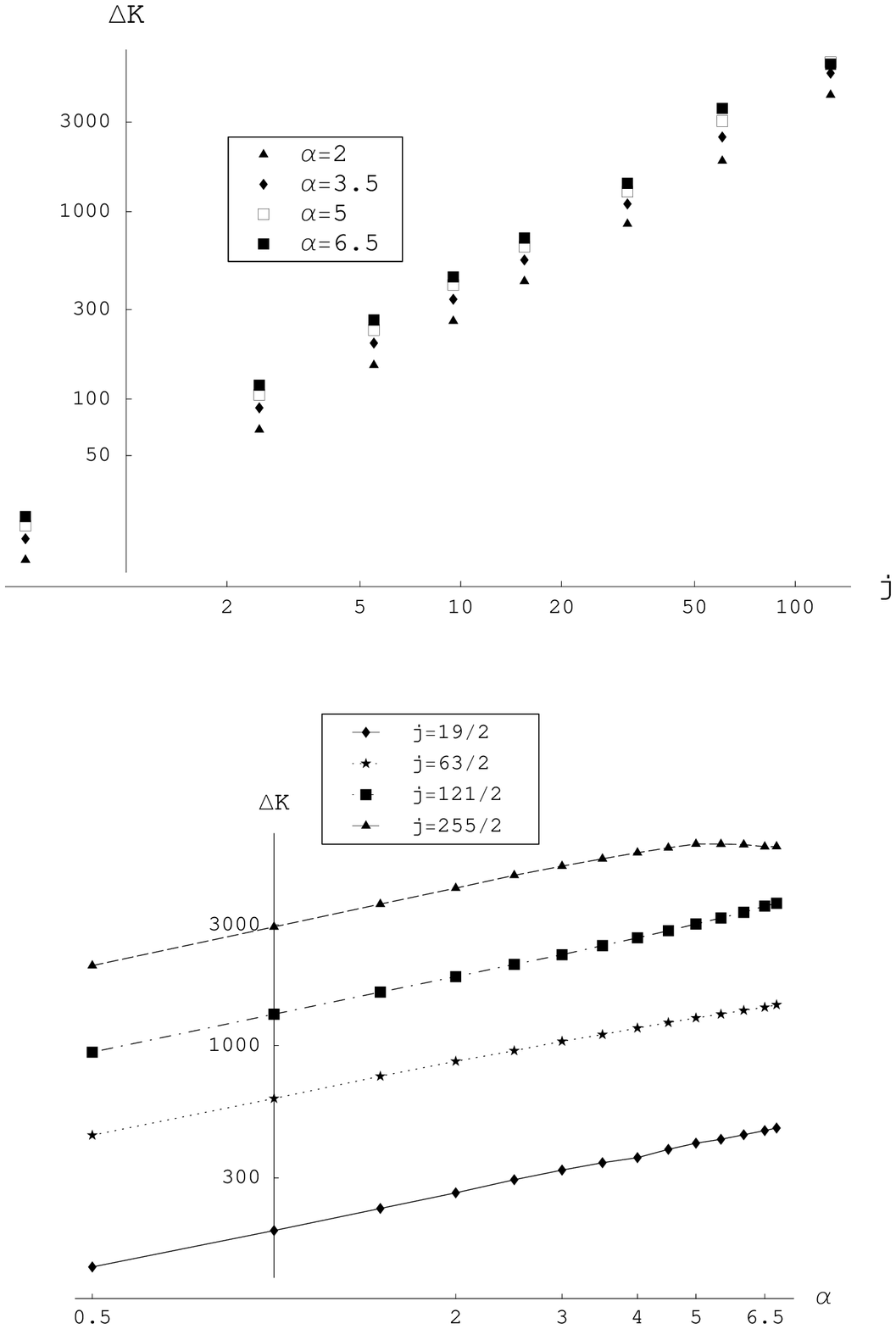}
\caption{The widths of curvature distributions $\Delta K\equiv
\langle \Delta^2 K\rangle $ as function of $j$ (a) and $\alpha$ (b).
The scaling laws $\Delta K\sim j^{\nu}$ and $\Delta K \sim \alpha
^{\nu}$ are seen. }
\end{figure}
For comparison with the level clustering (dimerization) found in
one-boson spin models (\cite{Cibils:1995,Steib:1998}) it is
interesting to demonstrate the numerical evidence of the
trimerization in the $E\otimes e$ model. From Fig. 8 it is evident
that the clustering of levels with two and three dominant spacings
at fixed angular momenta causes notable deviations from both
Poisson and Wigner distributions. The third (localized) peak is
characteristic for the two-boson case of a two-level model, while
the one-boson case is distinguished by the two peak structure.
\begin{figure}
\includegraphics[width=0.8\hsize]
{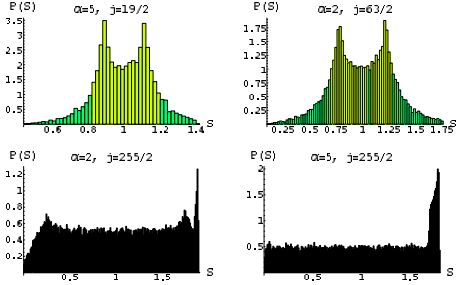} \caption{Level spacing distributions for particular
$j$. Dimerization clustering related to two dominating level
spacings is evident for $\alpha=2$ and $j=63/2$; Third peak from
localized states occurs at large $j$ and $\alpha$, dominating close
to the semiclassical limit (compare with Fig. 4). The crossover from
the linear dimerized regime to the nonlinear localized regime with
increasing $j$ and $\alpha$ is evident.}
\end{figure}
The long-range statistics comprises a characteristics of the
spectrum defined as its "rigidity measure". Among possible forms
of this parameter one can name the fluctuation of level number
$\delta ^2 N$ in an energy band of the widths $E$, the $\Delta_3$
statistics \cite{Mehta:1960,Evangelou:00,Shklovskii:93} etc. The
RMT, in its turn, predicts a characteristic behaviour for all
these measures which should scale as $\log E$ (or $\log \langle N
\rangle$), meanwhile for fully uncorrelated sequences of levels
the linear scaling $\sim \langle N \rangle$ is expected. In Fig.8
the sample example of the long-range statistics $\langle \Delta_3
\rangle$ is given. It is seen that the behaviour of the spectral
rigidity of the JT systems conforms rather to the RMT patterns.

\begin{figure}[h]
\includegraphics[width=0.8\hsize]
{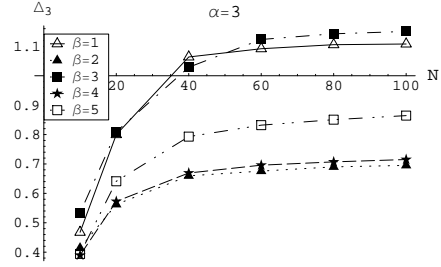} \caption{Example of the $\Delta_3$-statistics
(spectral rigidity) for the JT model. The long-range statistics
appears to show the log-like behaviour as function of $\langle N
\rangle$ typical for the RMT predictions (fully developed quantum
chaos).}
\end{figure}


\section{Vibronic spectra of $E\otimes e$ Jahn-Teller model as
 a Calogero-Moser gas of pseudo-particles with repulsion.}
 The approach of the present section is to some extent complementary to the statistical
 approach of the preceding section.
  The mapping of energy spectra $E_n(\alpha )$ of non-integrable
Hamiltonians with one parameter of non-integrability ($\alpha$),
$H=H_0+\alpha V$ where $H_0$ is integrable, on a classical
integrable Calogero-Moser model of interacting pseudo-particles
moving in a pseudo-time $\alpha$ is a very useful tool for
investigation of energy spectra of non-integrable systems first
introduced by Pechukas and Yukawa \cite{Pechukas:1983}. The
mapping is based on definition of the dynamic variables
$E_n(\alpha)\equiv x_n(\tau)$, $dx_n/d\tau= p_n(\tau)$,
 with the pseudo-time $\tau\equiv\alpha$.
Respective classical dynamic equations for pseudo-particles evolving
in the pseudo-time $\tau$ with repulsive interactions can be
rewritten for fluctuations $\delta_{2n}$ defined as
$x_{2n+1}-x_{2n}=1+\delta_{2n}$.
The Calogero-Moser set of equations reads
\begin{eqnarray}
\frac{dp_n}{d\tau}=2\sum_{m(\neq
n)}\frac{L_{nm}L_{mn}}{(x_m-x_n)^3}\nonumber\\
\frac{dL_{mn}}{d\tau}=\sum_{l\neq
(m,n)}L_{ml}L_{ln}\left[\frac{1}{(x_n-x_l)^2}-\frac{1}{(x_m-x_l)^2}\right
], \label{C1}
\end{eqnarray}
where
\begin{eqnarray}
L_{mn}(\tau)& =&(x_n(\tau)- x_m(\tau))\cdot V_{mn}=-L_{nm}
\nonumber\\p_n &\equiv &\langle n(\tau)|V|n(\tau)\rangle\equiv
V_{nn} \,.
\end{eqnarray}
 $V_{mn}$ are matrix elements of the Hamiltonian. Special case is
the Hamiltonian with excited spectrum consisting of $2\times 2$
diagonal  clusters with diagonal and nearest neighbor off-diagonal
elements, $f_{n_rn_r}\equiv={f_d}_{n_r}$ and $f_{n_r+1,n_r}\equiv
{f_o}_{n_r+1}$, respectively.
For the case (\ref{28}) we have $V_{2 n+1, 2n}= {f_d}_{n_r}, \
V_{2n, 2n-1}= {f_o}_{n_r}$. There is
\begin{eqnarray}
 L_{2n+1 2n}= {f_d}_{n_r} (x_{2n+1}(\tau)-x_{2n}(\tau)), \nonumber\\
{f_d}_{n_r} = 2  \sqrt 2 \sqrt{n+1+|j-1/2|},\nonumber\\
L_{2n-1 2n}= {f_o}_{n_r}(x_{2n}(\tau)-x_{2n-1}(\tau)),
\nonumber\\
{f_o}_{n_r} =-\sqrt 2 \sqrt{n}\,. \label{C2}
\end{eqnarray}
For what follows we define small fluctuation $\delta $ by
$x_{2n+1}-x_{2n}\equiv 1+\delta_{2n}$, and approximate
$(x_{2n+1}-x_{2n})^{-1}\approx
1-\delta_{2n}+\delta_{2n}^2-\delta_{2n}^3$. Then, for energy
fluctuations $\delta_{2n}$, $\delta_{2n+1}$ from (\ref{28}),
(\ref{C1}) and (\ref{C2}) we get nonlinear equations with the
lowest powers of nonlinearity in the form
\begin{eqnarray}
\frac{\partial^2\bar \delta_{2n+1}}{\partial \tau^2}-4{f_d}_{n_r+1}
^2 (\bar \delta
_{2n+2}+\bar\delta_{2n}-2\bar\delta_{2n+1})\nonumber\\
=8({f_d}_{n_r+1} ^2-{f_o}_{n_r+1}^2) \bar\delta_{2n+1} - \nonumber
\\4 ({f_d}_{n_r+1} ^2-{f_d}_{n_r}^2) \bar\delta_{2n}
\nonumber\\
-8 {f_o}_{n_r+1}^2\bar\delta_{2n+1}^2+4 {f_d}_{n_r+1}^2
\bar\delta_{2n+2}^2+4{f_d}_{n_r}^2\bar\delta_{2n}^2 + \nonumber \\
+O(\bar\delta^3) \,,\label{6}
\end{eqnarray}
\begin{eqnarray}
\frac{\partial^2\bar \delta_{2n}}{\partial \tau^2}-4{f_o}_{n_r+1}^2
(\bar \delta
_{2n+1}+\bar\delta_{2n-1}-2\bar\delta_{2n})\nonumber\\
=8 ({f_o}_{n_r+1}^2-{f_d}_{n_r+1}^2) \bar\delta_{2n} - \nonumber
\\ 4 ({f_o}_{n_r+1}^2-{f_o}_{n_r}^2)
\bar\delta_{2n-1}\nonumber\\
-8 {f_d}_{n_r}^2\bar\delta_{2n}^2+4 {f_o}_{n_r+1}^2
\bar\delta_{2n+1}^2+4{f_o}_{n_r}^2\bar\delta_{2n-1}^2 \nonumber \\
+O(\bar\delta^3) . \label{7}
\end{eqnarray}
 Eqs. (\ref{6}) and (\ref{7}) can
be approximated by  taking second terms on the l.h.s. as second
derivatives. In view of definitions (\ref{C2}) for large $j\gg n$
we have $|{f_o}_{n_r+1}< |{f_d}_{n_r}|, \ f_d(n_r+1) \approx
{f_d}_{n_r}$. If we redefine $\bar\delta_{2n}= \delta_{2n}-1$ and
assume for the deviations plausible relations
$\bar\delta_{2n}^2\approx\bar\delta_{2n+2}^2\approx\bar\delta_{2n+1}^2$
in Eq. (\ref{6}) and (\ref{7}), we obtain, up to the second order
\begin{eqnarray}
\frac{\partial^2\bar \delta_{2n+1}}{\partial
\tau^2}-4{f_d}_{n_r+1} ^2 \frac{\partial^2\bar
\delta_{2n+1}}{\partial n^2}\approx \nonumber \\
8 {f_d}_{n_r+1}
^2 (\bar\delta_{2n+1}+\bar\delta_{2n+1}^2) ,
 \label{6a}
\end{eqnarray}
\begin{eqnarray} \frac{\partial^2\bar
\delta_{2n}}{\partial \tau^2}-4{f_o}_{n_r+1}^2
\frac{\partial^2\bar \delta_{2n}}{\partial n^2}\approx \nonumber
\\
-8{f_d}_{n_r}^2(\bar\delta_{2n}+\bar\delta_{2n}^2). \label{7a}
\end{eqnarray}
 In view of different signs of the potentials on the r.h.s. of
equations (\ref{6a}) and (\ref{7a}) even and odd pseudo-particles
 can be related as moving in real and imaginary space, $\tau\rightarrow i\tau,\  n\rightarrow i n$,
 respectively.\\
Analogously, alternative set of equations reads
\begin{equation}
\frac{\partial^2\bar \delta_{2n+1}}{\partial
\tau^2}-4{f_o}_{n_r+1}^2 \frac{\partial^2\bar
\delta_{2n+1}}{\partial n^2}\approx 8 {f_d}_{n_r+1}^2
(\bar\delta_{2n} +\bar\delta_{2n}^2) , \label{11b}
\end{equation}
\begin{equation}
\frac{\partial^2\bar \delta_{2n}}{\partial \tau^2}-4{f_d}_{n_r}^2
\frac{\partial^2\bar \delta_{2n}}{\partial n^2}\approx -8
{f_d}_{n_r+1} ^2 (\bar\delta_{2n+1} +\bar\delta_{2n+1}^2) .
\label{12b}
\end{equation}
Transition to the imaginary space changes the sign of the potential,
i.e. the particle turns to the tunneling domain and the
pseudo-particles change their parity, e.g.
$\bar\delta_{2n+1}(v_1\equiv 2 {f_d}_{n_r}\rightarrow
\bar\delta_{2n}(v_1)$, $\bar\delta_{2n}(v_2\equiv 2{f_d}_{n_r}
\rightarrow \bar\delta_{2n+1}(v_2\equiv 2 {f_o}_{n_r+1})$, so that
the trajectories interchange their velocities,
$\bar\delta_{2n}(v_2)\equiv 2{f_o}_{n_r+1}\rightarrow
\bar\delta_{2n}(v_1)\equiv 2 {f_d}_{n_r} $ and
$\bar\delta_{2n+1}(v_1\equiv 2{f_d}_{n_r} \rightarrow
\bar\delta_{2n+1}(v_2)\equiv 2 {f_o}_{n_r+1}$. By another words, the
particles interchange their velocities, $v_1, v_2$ when transferring
from real to imaginary space and vice versa. This scenario can be
understood as series of soliton collisions related to two subsequent
levels.\\
Indeed, equation (\ref{6a}) for large $j$ (${f_d}_{n_r+1} \sim
2\sqrt{|2j-1|}$) admits solution in form of the propagating
nonlinear pulse $\bar\delta (\zeta=n-v\tau)$
\begin{equation}
\bar\delta_{2n+1} (\zeta-\zeta_0)=-\frac{3}{2}\cosh^{-2}\left (
\frac{\zeta-\zeta_0}{L}\right ),
\end{equation}
where $\zeta_0=n_0-v\tau_0$ restores the translational invariance
in the system of levels and $L=
(v^2-({f_d}_{n_r+1})^2)^{1/2}/[2\sqrt 2|{f_d}_{n_r+1}|]$ is the
soliton width. The corresponding coordinate is identified from
$\partial x_{2n+1}/\partial n =1+\delta_{2n+1}$ as a soliton
(kink) solution
\begin{equation}
x_{2n+1}=2n-\frac{3L}{2}\tanh \left (\frac{\zeta-\zeta_0}{L}\right).
\end{equation}
If in equations (\ref{6a}) and (\ref{7a}) one approximates
$\bar\delta_{2n+2}\approx\bar\delta _{2n}$ and assumes
$\bar\delta_{2n+1}=-\bar\delta _{2n}$ one obtains approximately
\begin{eqnarray}
\delta_{2n+1}\approx -
\frac{{f_d}_{n_r+1}^2+{f_d}_{n_r}^2}{2{f_o}_{n_r+1}^2}\delta_{2n}(1-\delta_{2n}).\label{10}
\end{eqnarray}
Equation (\ref{10}) is a well known logistic equation implying
transition to the chaotic regime for $A=\frac{{f_d}_{n_r+1}
^2+{f_d}_{n_r} ^2}{2{f_o}_{n_r+1}^2}\sim \frac{n+|j-1/2|}{n}\geq
A_{crit}=3.56994 \dots$.\\
 Conventionally, the term ``quantum chaos'' is used to denote
the traces of {\it classical} chaotic behavior at a quantum level.
As already mentioned in the Introduction, the classical counterpart
of the system under consideration cannot be defined uniquely. The
semiclassical approximation in two-level systems generically leads
to classical chaotic patterns as a result of nonlinear coupling
between two subsystems, boson and electron, considered respectively
as classical and quantum; The onset of classical chaos corresponds
to the energies above the first diabatic line. It is to be
emphasized that the mentioned chaotic behavior refers to purely
quantum regime of medium $j$ and $n_r$ between the weak coupling
with $j\ll n_r$ (dimerized pairs of oscillators) and strong coupling
with $j\gg n_r$ (kink lattice) domains. Thus, this chaotic behavior
can be regarded as being of essentially quantum nature. One can
conclude, that the mapping of the quantum system on the classical
Calogero-Moser gas with repulsive interactions enables one to use
the classical formalism for describing the system via its quantum
numbers. So the remarkable feature of this approach is its ability
to represent quantum chaos by classical equations.
 \section{Acknowledgments}
 \label{ack} The support of the project by the Grant Agency of the Czech Republic 
 No. 202/06/0396 is greatly acknowledged. Partial support is acknowledged also from
the project No.  2/6073/26 of the Grant Agency VEGA, Bratislava.
%






\end{document}